\documentclass{cjaa}                   

\usepackage{graphicx}                   
\input{epsf.sty}                        
\input{psfig.sty}                       

\begin{document}

   \title{Study the X-ray Dust Scattering Halos with Chandra Observations of Cygnus X-3 and Cygnus X-1}

   \volnopage{Vol.0 (200x) No.0, 000--000}      
   \setcounter{page}{1}          

   \author{Jingen Xiang
      \inst{1}\mailto{}
   \and Shuang Nan Zhang
      \inst{1,2}
   \and Yangsen Yao
      \inst{3}
      }
   \offprints{Jin-Gen Xiang}                   

   \institute{Physics Department and Center for Astrophysics, Tsinghua University, Beijing 100084, China\\
             \email{xjg01@mails.tsinghua.edu.cn}
        \and
             Department of Physics, University of Alabama in Huntsville, Huntsville, AL 35899\\
        \and
             Department of Astronomy, University Massachusetts, Amherst, MA 01003\\
          }

   \date{Received~~2004 month day; accepted~~200x~~month day}

   \abstract{We improve the method proposed by Yao \emph{et al} (2003)
   to resolve the X-ray dust scattering halos of point
sources. Using this method we re-analyze the Cygnus X-1 data
observed with {\it Chandra} (ObsID 1511) and derive the halo
radial profile in different energy bands and the fractional halo
intensity (FHI) as $I(E)=0.402\times E_{{\rm keV}}^{-2}$. We also
apply the method to the Cygnus X-3 data ({\it Chandra} ObsID 425)
and derive the halo radial profile from the first order data with
the {\it Chandra} ACIS+HETG. It is found that the halo radial
profile could be fit by the halo model MRN (Mathis, Rumpl $\&$
Nordsieck, 1977) and WD01 (Weingartner $\&$ Draine, 2001); the
dust clouds should be located at between 1/2 to 1 of the distance
to Cygnus X-1 and between 1/6 to 3/4 (from MRN model) or 1/6 to
2/3 (from WD01 model) of the distance to Cygnus X-3, respectively.
    }
   \authorrunning{J. G. Xiang, S. N. Zhang \& Y. S. Yao}            
   \titlerunning{The Halos of Cygnus X-3 and Cygnus X-1 }  

   \maketitle

%
%
\section{Introduction}           
\label{sect:intro} X-ray halos are formed by the small-angle
scatterings of X-rays off dust grains in the interstellar medium.
The scatterings are significantly affected by (a) the energy of
radiation; (b) the optical depth of the scattering, due to the
effects of multiple scatterings; (c) the grain size distribution
and compositions; and (d) the spatial distribution of dusts along
the line of sight (Mathis $\&$ Lee, 1991, Mathis, Rumpl $\&$
Nordsieck, 1977). Analyzing the properties of  X-ray halos is an
important tool to study the interstellar grains, which plays a
central role in the astrophysical study of the interstellar
medium, such as the thermodynamics and chemistry
of the gas and the dynamics of star formation.\\

Before {\it Chandra} was lunched, it has been very difficult to
get the accurate physical parameters of the X-rays halos due to
the poor angular resolution of the previous instruments. With
excellent angular resolution, good energy resolution and broad
energy band, the {\it Chandra} ACIS
\footnote{http://cxc.harvard.edu/proposer/POG/html/ACIS.html} is
so far the best instrument for studying the X-ray halos. However,
the direct images of bright sources obtained with ACIS usually
suffer from severe pile-up. Although the data in CC-mode or HETGs
\footnote{http://space.mit.edu/HETG/index.html} have either no or
less serious pileup, the data in CC mode have only one dimensional
images and the data in HETGs are mixed with different energies and
radius data, from which we can not get the halo's radial profile
directly. By making use of the assumption that the real halo
should be an isotropic image, we have reported the reconstruction
of the images of X-ray halos from the data obtained with the HETGS
and/or in CC mode. The detailed method has been described in Yao
\emph{et al.} (2003, here after Paper I).

In this paper, we improve the method to resolve the X-ray dust
scattering halos of point sources which is proposed in Paper I.
The method can resolve the halo more accurately than the method in
Paper I, even when some data are contaminated. Furthermore, we
modify the method to create the PSF of {\it Chandra}. These will
be shown in section 2. Using this method we reanalyze the Cygnus
X-1 data and also apply the new method to the Cygnus X-3 data,
which will be shown in section 3. Finally we give our conclusions
and make some discussions in section 4.


\section{The method, point spread function and simulation}
\label{sect:method}

In Paper I, we have reported that if the flux of a point source
plus its X-ray halo is isotropically distributed and centered at
the point source as $F(r)$, and the projection process in which
the two-dimensions halo image is projected to one dimension image
can be represented by a matrix operator $M(r,d)$, then the
projected flux distribution $P(d)$ is
\begin{equation}
P(d)=M(r,d)\times F(r),
\end{equation}
where $r$ is the distance from the centroid source position and
$d$ is the distance from the projection center (refer ro Fig. 1).
After calculating the inverse matrix of the operator $M(r,d)$, we
can get the source flux $F(r)$. In CC-mode, we can only get the
count rate $C(d)$, but not the flux projection $P(d)$ directly.
With the exposure map of CCDs calculated, we can get another
equation
\begin{equation}
{\rm{exposure}\ \rm{map} \choose \rm{matrix}} \times M(r,d)\times
F(r) = M'(r,d) \times F(r) = C(d),
\end{equation}
where $M'(r,d)={\rm{exposure}\ \rm{map} \choose \rm{matrix}}
\times M(r,d)$ is another matrix. Because the inverse matrix of
the operator $M'(r,d)$ may not exist, we use the iterative method
to solve the above equations. Using the iterative method, we can
get the result even though some data in $P(d)$ or $C(d)$ are
imperfect or defected. For example, the large angle data
$C(d^{*})$ in CC-mode zeroth order are usually contaminated by the
data from the HETGs 1st order. In this case we can set these data
$C(d^{*})$ and the corresponding matrix elements of $M(r,d^{*})$
to zero. In the process of reconstruction, we can set some limits
based on physical constraints, for example the flux of halo should
be larger than zero and the flux in the small angle should be
larger than the one in the large angle.
\begin{figure}
  \begin{center}
   \hspace{3mm}\psfig{figure=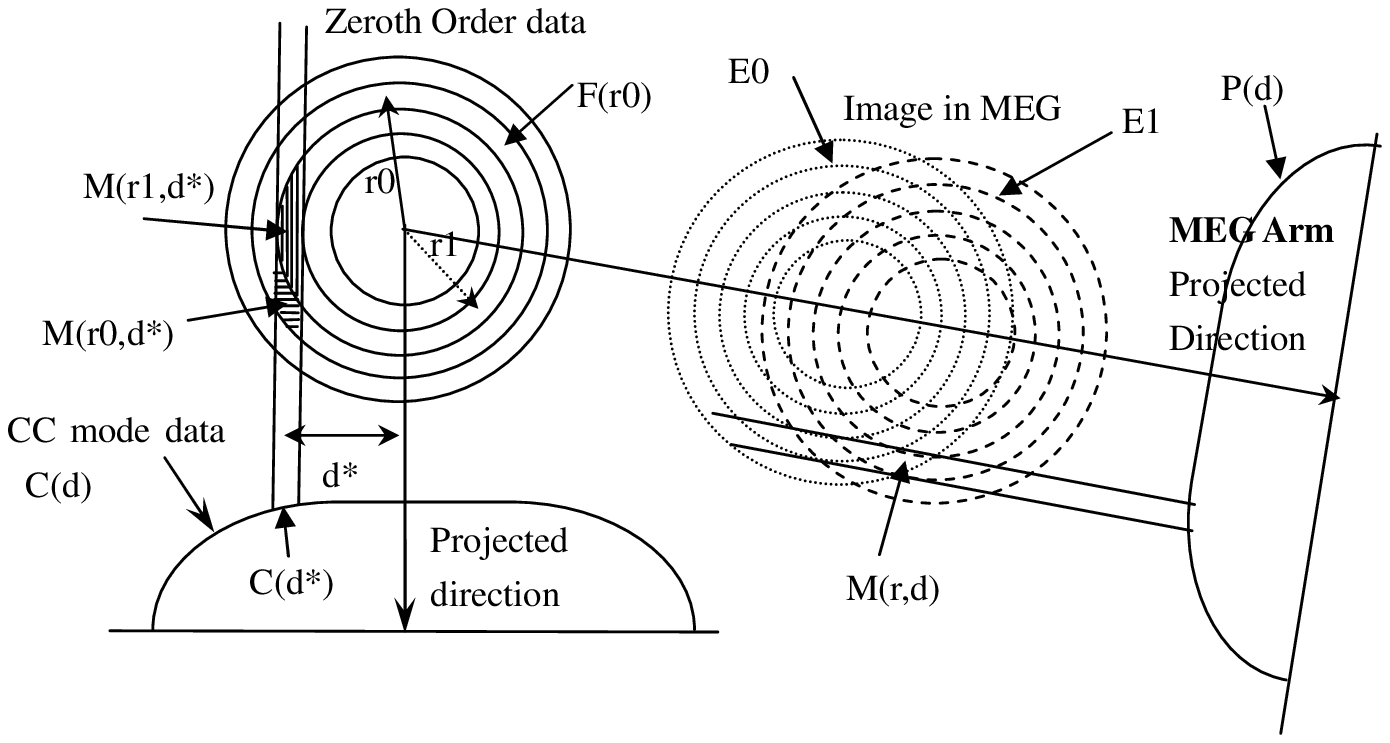,width=120mm,height=64mm,angle=0.0}
    \vspace{-5mm}
\caption{The projection of the photons in zeroth order image along
the read-out direction and the projection of the photons along the
grating arm.}
   \label{Fig:lightcurve-ADAri}
   \end{center}
\end{figure}

We use the steepest descent method (Marcos Raydan, \emph{et al},
2001) to solve equation 3. Then the iterative process can be
expressed as
\begin{eqnarray}
F^{(k+1)}(r)=F^{(k)}(r)&+&\frac{[C(d)-M'(r,d)F^{(k)}(r)]^T[C(d)-M'(r,d)F^{(k)}(r)]}{[M'(r,d)[C(d)-M'(r,d)F^{(k)}(r)]]^T[C(d)-M'(r,d)F^{(k)}(r)]}
\nonumber\\&\times& [C(d)-M'(r,d)F^{(k)}(r)]
\end{eqnarray}
and where $F^{(k+1)}(r)$ and $F^{(k)}(r)$ are the values of F(r)
in the ($k+1$)th and $k$th iterative loops,
$[C(d)-M'(r,d)F^{(k)}(r)]^{T}$ is the transpose of the matrix
$C(d)-M'(r,d)F^{(k)}(r)$. In our iterative process, the loop is
stopped when
$\frac{1}{N}\sum_{d=1}^{N}(\frac{M'(r,d)F(r)-C(d)}{\Delta C(d)})^2
< 0.05$, where $N$ is number of $C(d)$ and $\Delta C(d)$ is the
error of $C(d)$.

The accurate {\it Chandra} PSF (Point Spread Function) is
important for reconstructing the halo accurately. We used the {\it
MARX} simulator \footnote{http://space.mit.edu/ASC/MARX/} to
calculate the PSF, but found that at large angles (above 50
arcsec), the PSF from the MARX simulation is very different from
the radial flux profile of a point source without halo. Due to the
pile-up for bright point sources in the center, we cannot get the
PSF from the observation of a bright point source at small angles
(below 3--5 arcsec). Therefore we calculated the PSF from the
observation data of bright point sources without halo (for example
Her X-1) at large angles and from the MARX simulation at small
angles.

 To test our method, we produced with MARX 4.0 simulator a point source plus one disk
source to mimic a point source with its X-ray halo observed with
{\it Chandra} ACIS+HETGs in TE mode. The reconstructed halo radial
surface brightness distribution is consistent with the simulation
input, and the recovered FHI (Fractional Halo Intensity) (49.25\%)
is also consistent with the input value (50\%). We thus conclude
that the improved method can resolve the radial profile of the
halo accurately.
\section{Application to Cygnus X-3 and Cygnus X-1}
\label{sect:application} Cygnus X-3 is an X-ray binary with more
than 40$\%$ halo flux in 0.1--2.4 keV ({\it ROSAT} energy range)
(Predelh \emph{et al}, 1995). The bright X-ray halo has been used
to determine the distance of the source (Predelh, \emph{et al},
2000). Cygnus X-3 is so bright that there was severe piled-up in
the zeroth order image of {\it Chandra} ACIS. We use our method to
resolve the halo from the data with the highest statistical
quality, which was observed on 2000 April 4 (ObsID 425) with {\it
Chandra} ACIS+HETGs in TE mode.

We use the first order data (HEG $\pm 1$ and MEG $\pm 1$) within
60 arcsec around the source position. First, we filter the data in
selected regions using the CCD energy measurement along the
grating arm and generate the exposure map for each energy band and
 region. Dividing the count images by the exposure map, an image
in flux units is produced. Then we project the image along the
grating arm and get the projected flux in units of photons
cm$^{-2}$ s$^{-1}$ arcsec$^{-2}$ per energy band. Finally, we sum
the data from the four regions (HEG $\pm 1$ and MEG $\pm 1$) and
derive the projected total flux. With the same procedure, we get
the projections of PSFs from other sources with negligible halos
observed with {\it Chandra} ACIS+HETG in TE mode. In order to
improve the statistical quality, we use data from four
observations, namely, Her X-1 (ObsID 2749) and PKS 2155-304 (ObsID
337, 3167 and 3706). After extracting the projections of PSF, we
obtain the pure halo projections. Finally we calculate the
operator matrix and derive the halo radial profile in units of
photons cm$^{-2}$ s$^{-1}$ arcsec$^{-2}$. The reconstructed halo
radial profile is shown in Fig. 2.
\begin{figure}
   \begin{center}
   \hspace{3mm}\psfig{figure=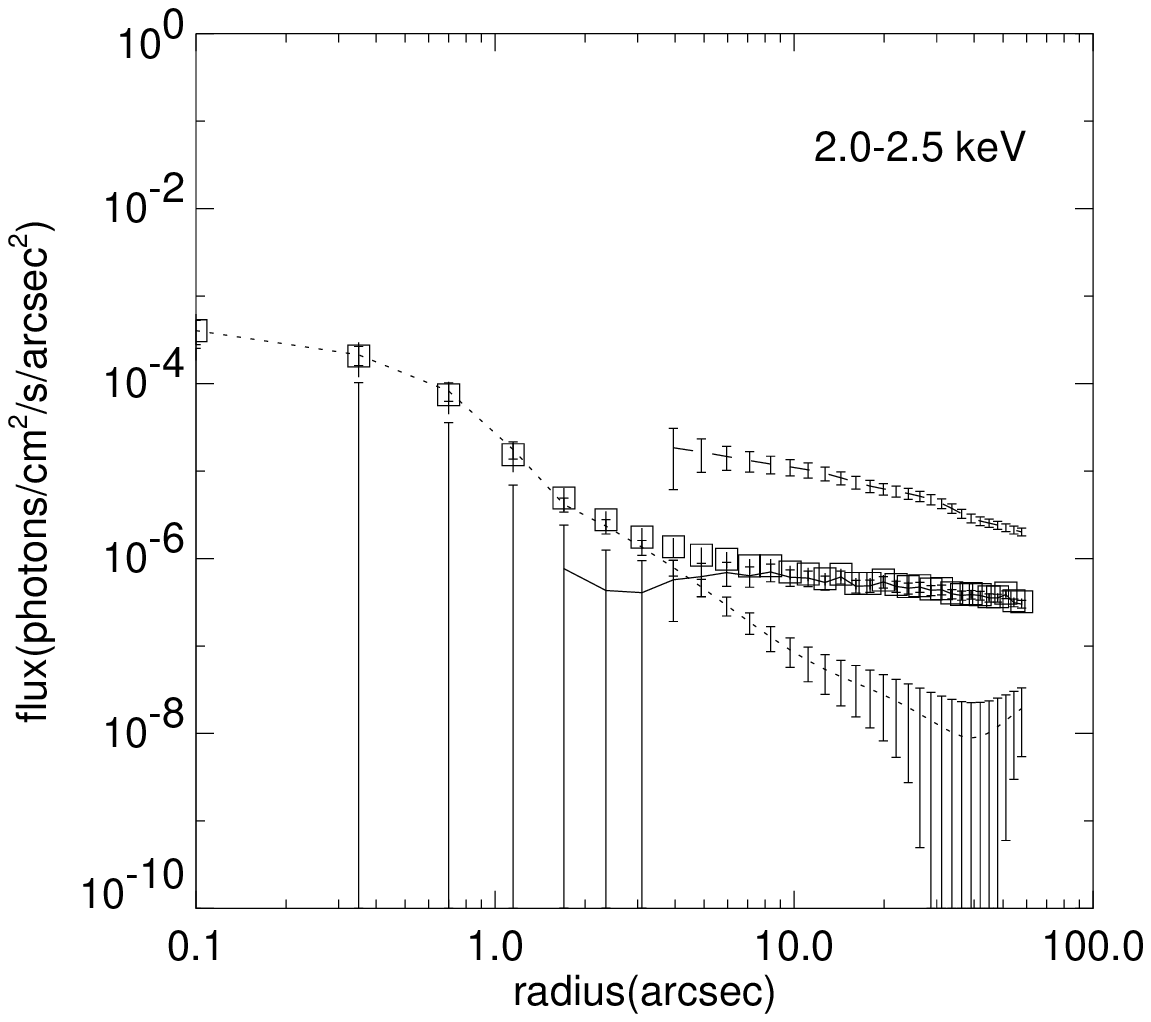,width=68mm,height=60mm,angle=0.0}
   \hspace{3mm}\psfig{figure=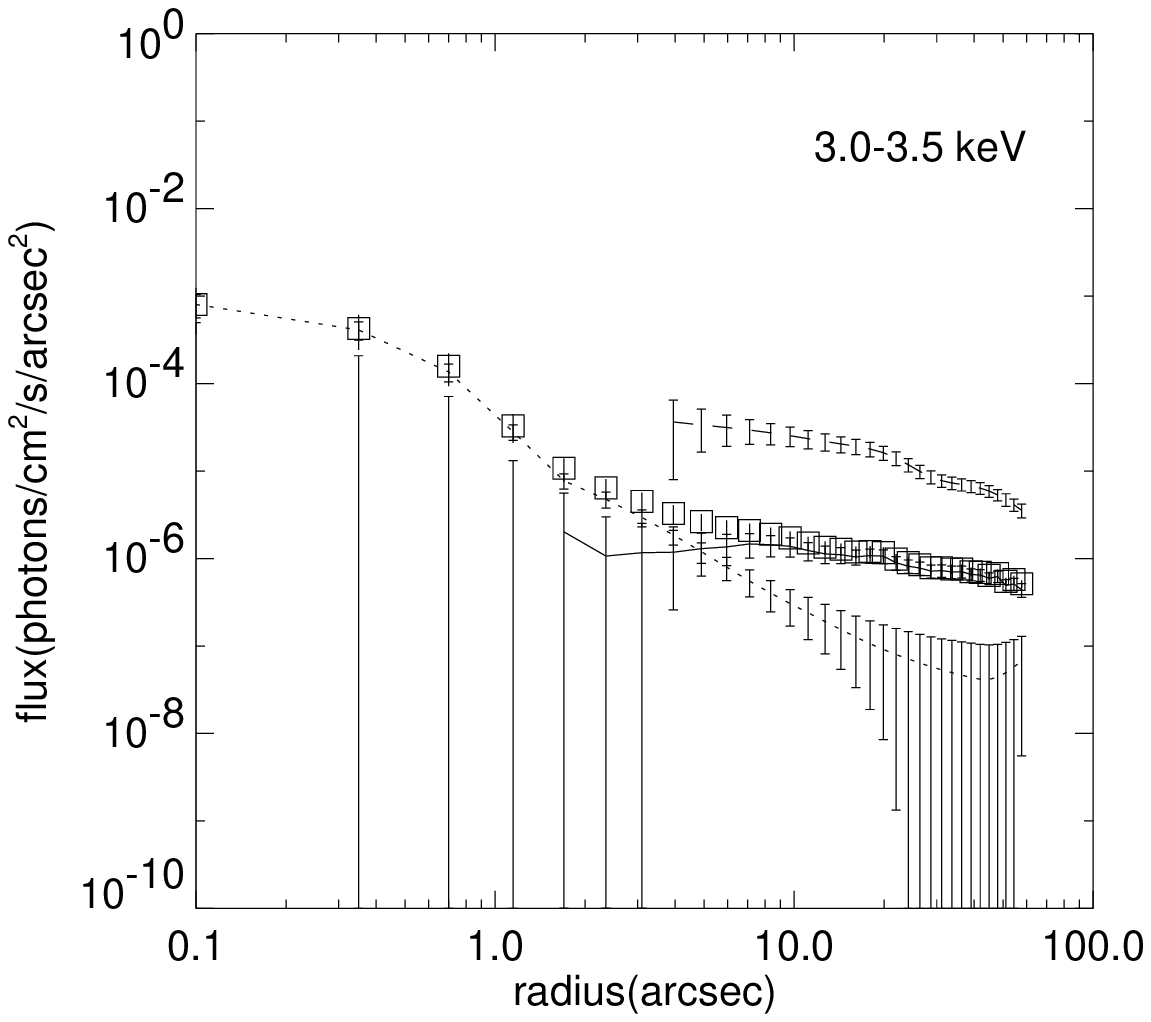,width=68mm,height=60mm,angle=0.0}
  \vspace{-5mm}
  \caption{The reconstructed halo radial profile of Cygnus X-3. The squares data are the
   1st order projection. The dotted line is the projection of PSF
   in 1st order, the solid line is the pure halo projection, and the dashed line
   is the reconstructed halo radial profile multiplied by 10 for clarity.
   Due to the poor statistical quality, the halo within 4 arcsec is not reconstructed.}
   \label{Fig:lightcurve-ADAri}
   \end{center}
\end{figure}
\begin{figure}
  \begin{minipage}[t]{0.5\linewidth}
  \centering
  \includegraphics[width=70mm,height=60mm]{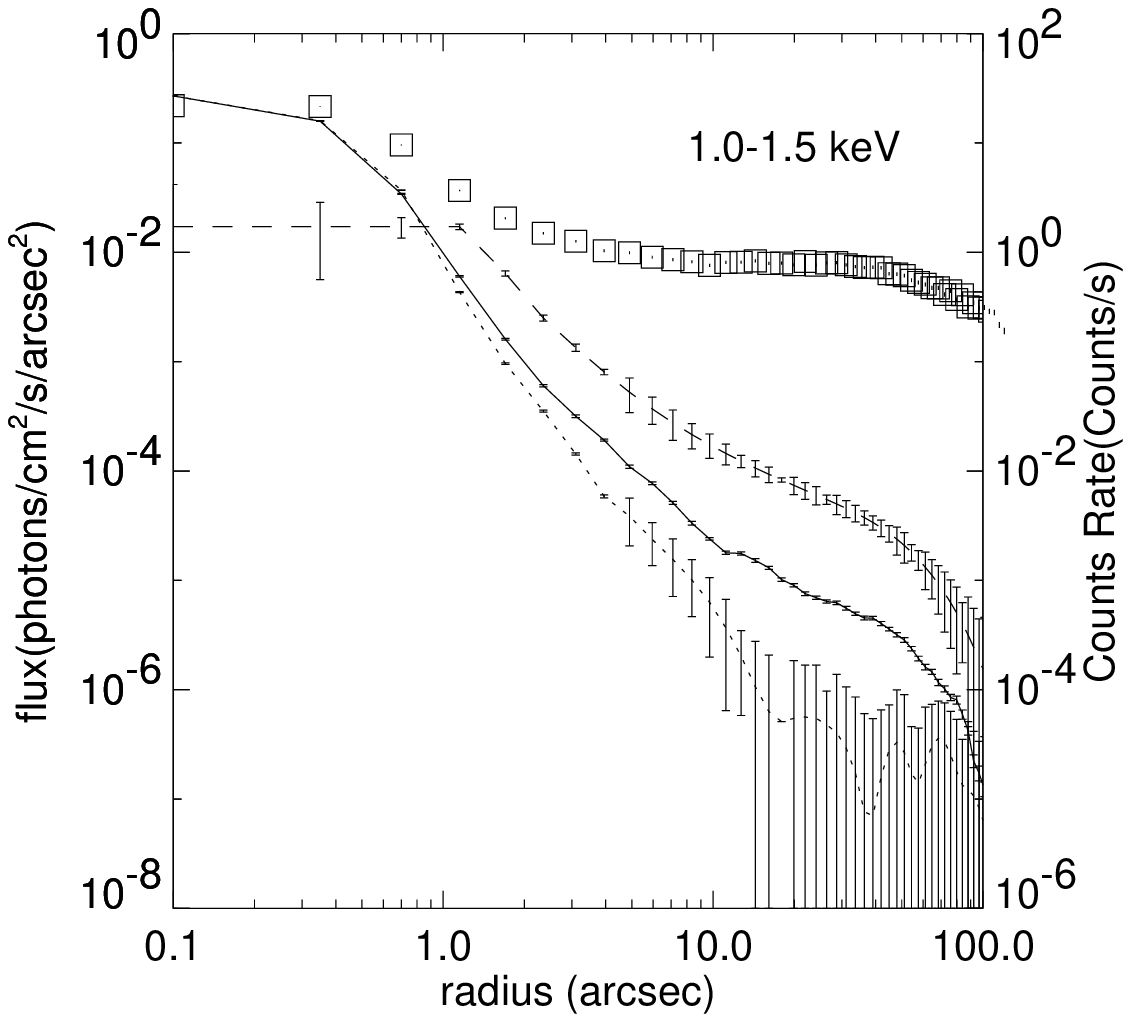}
  \vspace{-5mm}
  \caption{{The reconstructed halo radial profile of Cygnus X-1. The squares data are the projected flux disribution (counts/s). The solid line
  is reconstructed flux distribution (convolved by the PSF), the
   dotted line is the PSF from the MARX 4.0 simulation and the {\it Chandra} Her X-1 data
   (ObsID 2749), and the dashed line is the pure halo radial profile multiplied
   by 10 for clarity.} }
  \end{minipage}%
  \begin{minipage}[t]{0.5\textwidth}
  \centering
  \includegraphics[width=70mm,height=60mm]{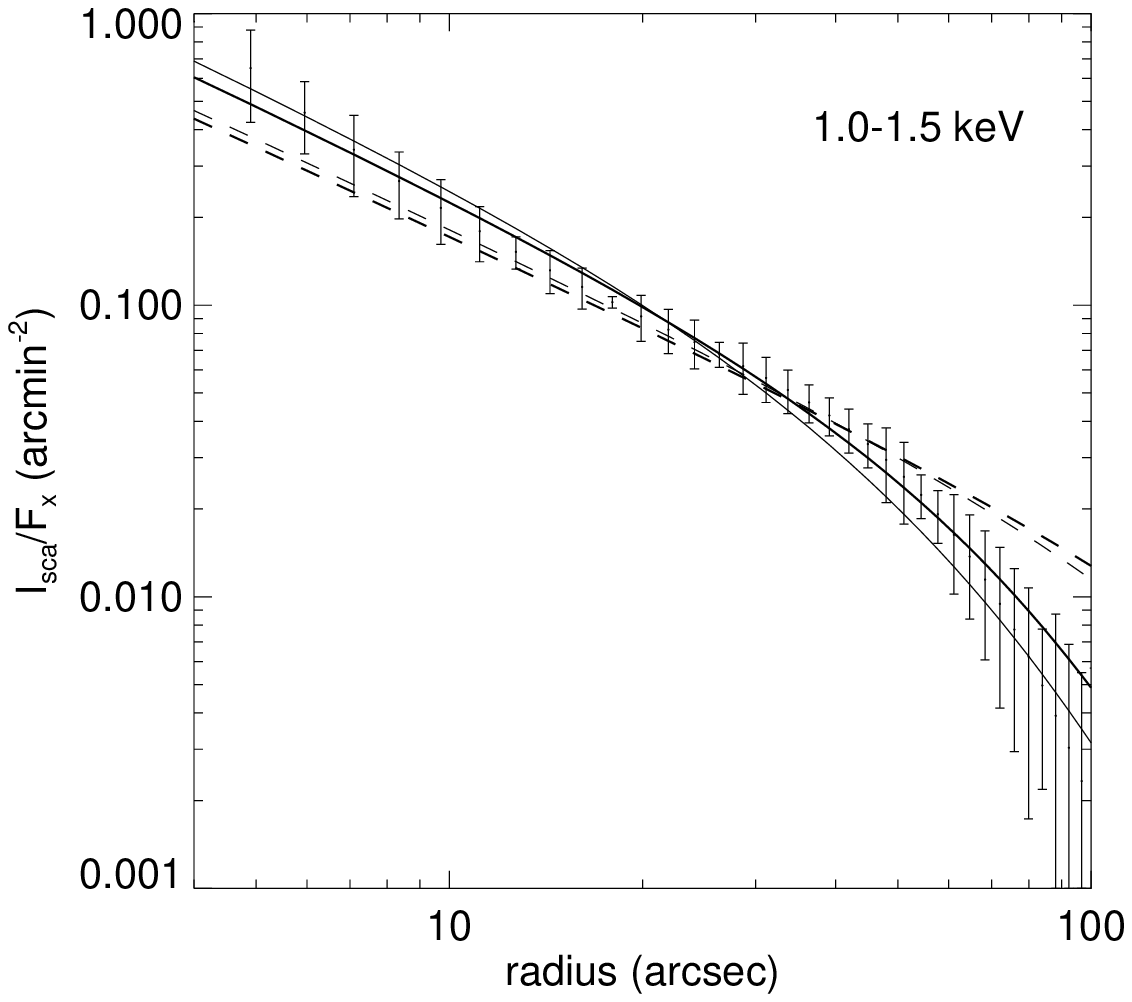}
  \vspace{-5mm}
  \caption{{The X-ray halo profile of Cygnus X-1, fitted using the halo models MRN and
     WD01. The solid lines are the models MRN (thick line) and WD01 (thin line) where the dust clouds are located between 1/2 to 1 of the
     distance to the source. The dashed lines are the models MRN (thick line) and WD01 (thin line) with smoothly
     distributed dust.}}
  \end{minipage}%
  \label{Fig:fig23}
\end{figure}

Cygnus X-1, discovered in 1965, is one of the brightest X-ray
sources in the sky. There is about 11\% halo flux in 0.1-2.4 keV
(Predehl \& Schmit, 1995). In Paper I, we tested our method on
{\it Chandra} data of Cygnus X-1. Here we re-analyze the data
observed on 2000 January 12 (ObsID 1511) with {\it Chandra}
ACIS+HETGs
in CC-mode with an effective exposure of 12.7 ks.\\
The zeroth order data are the projection C(d) in units of counts
arcsec$^{-2}$. Then we calculate the exposure map ${\rm{exposure}\
\rm{map} \choose \rm{matrix}}$ in units of counts cm$^{2}$ s
photon$^{-1}$ of the CCDs within $2^{'}$ from the source position.
Multiplying the exposure map matrix to the projected matrix
$M(r,d)$, we obtain the new matrix $M'(r,d)$. Using the iterative
method, we derive the radial flux $F(r)$ in units of photons
cm$^{-2}$ s$^{-1}$ arcsec$^{-2}$.

For energies above 4.0 keV, the grating arms extend into the $2'$
region and the zeroth order data in some bins are contaminated by
the grating events. Therefore we set these contaminated bins and
the corresponding operator matrix elements to zero. For example,
the data between 102 arcsec and 117 arcsec (corresponding to 30-32
bins) in energy region 4.5--5.0 keV and the corresponding matrix
elements $M[30:32][*]$ are set to zero. The reconstructed flux
distribution, the {\it Chandra} PSF and the extracted halo in
energy band 1.0--1.5 keV are shown in Fig. 3.

Then, we used the halo models MRN and WD01 to fit our halo radial
profile of Cygnus X-1 and Cygnus X-3. The two models have
different dust grain size distributions: $n(a)~a^{-3.5}$ for MRN
and $n(a)$ for WD01 is much more complex (refer to Weingartner
$\&$ Draine, 2001). For both sources we find the smoothly
distributed dust models cannot describe their halo profiles. Then
we change the position of the dust cloud manually in the models to
fit the halo profile. The best-fit results suggest that the dust
clouds should be located at between 1/2 to 1 of the distance to
Cygnus X-1 and between  1/6 to 3/4 (fitted by MRN) or 1/6 to 2/3
(fitted by WD01) of
the distance to Cygnus X-3, as shown in Fig. 4 and Fig. 5, respectively.\\

\begin{figure}
   \begin{center}
   \hspace{3mm}\psfig{figure=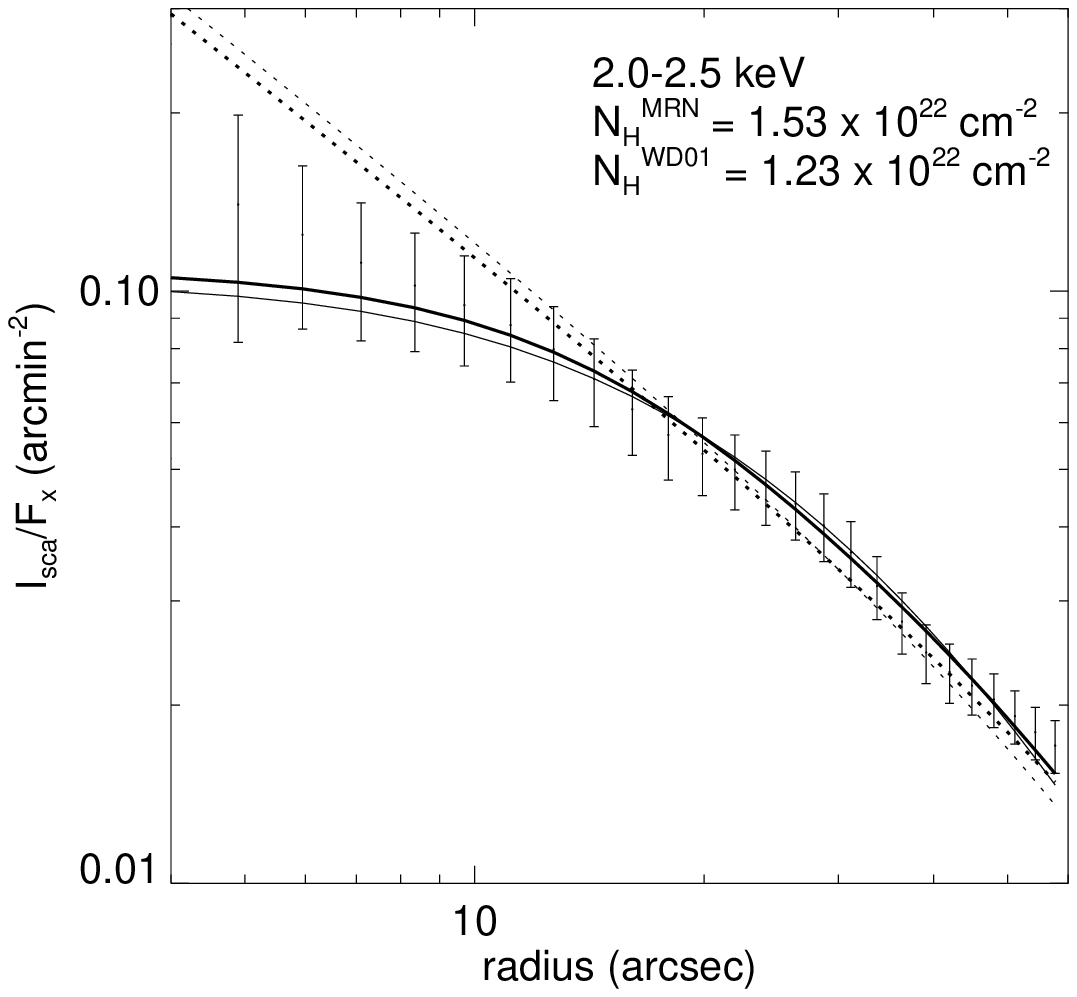,width=68mm,height=60mm,angle=0.0}
   \hspace{3mm}\psfig{figure=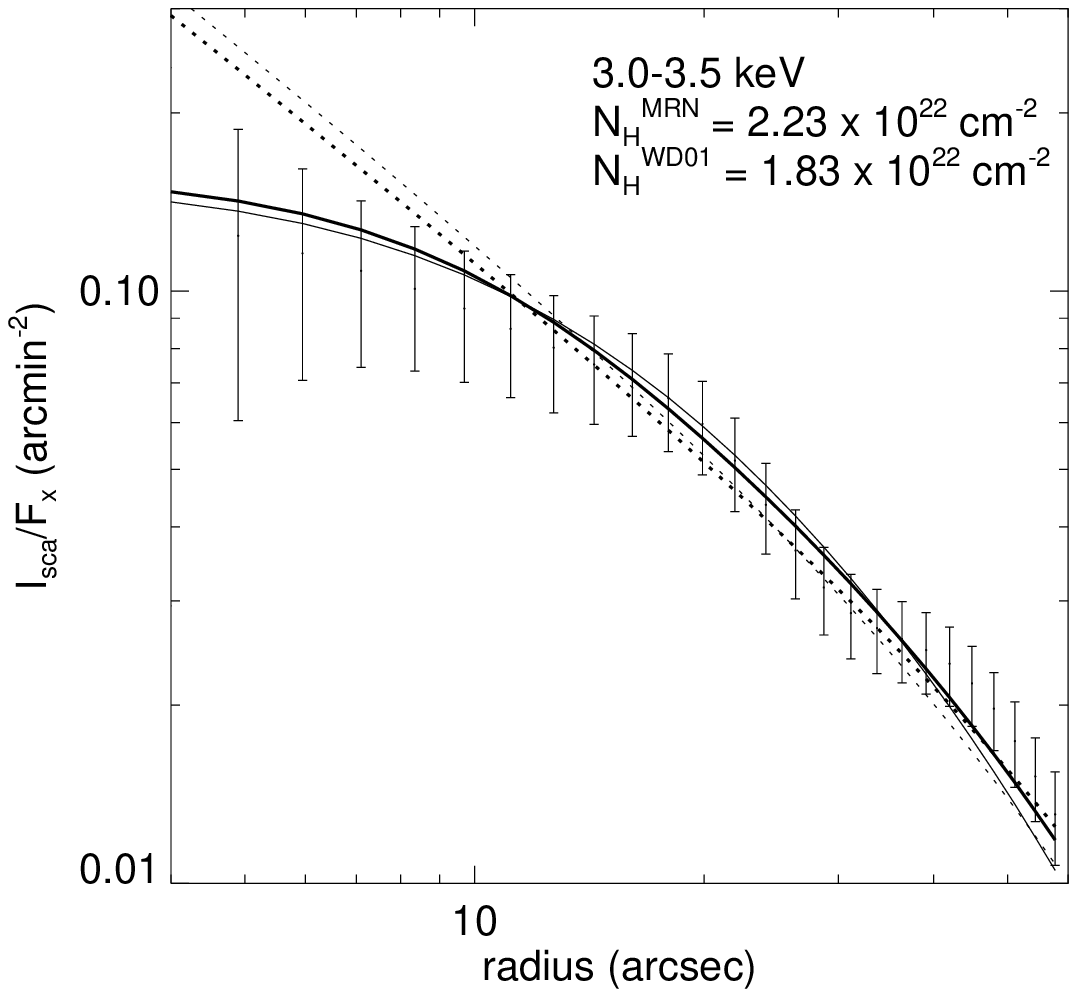,width=68mm,height=60mm,angle=0.0}
  \vspace{-5mm}
  \caption{X-ray halo profiles of Cygnus X-3 at 2.0--2.5 keV and
    3.0--3.5 keV fitted using the halo models MRN and WD01. The thick
    solid line is the MRN model where the dust clouds are located between
    1/6 to 3/4 of the distance to the source. The thin solid line is the WD01 model
    where the dust clouds are located between 1/6 to 2/3 of the distance to
    the source. The dotted lines are the model MRN (thick line) and WD01 (thin line)  with smoothly distributed dust.}
   \label{Fig:lightcurve-ADAri}
   \end{center}
\end{figure}
\begin{figure}
   \begin{center}
   \hspace{3mm}\psfig{figure=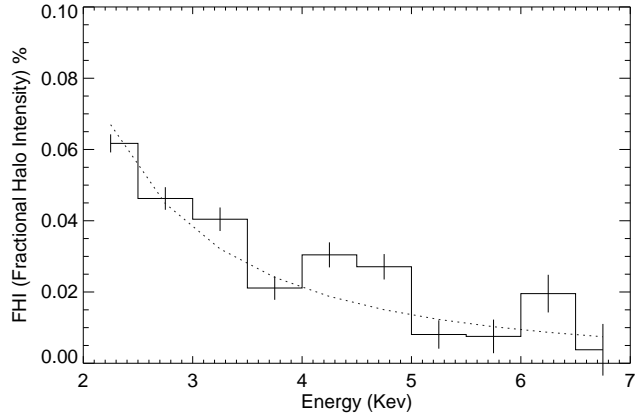,width=90mm,height=60mm,angle=0.0}
    \vspace{-5mm}
\caption{The FHI (relative to the source flux), as a function of
energy
   with the best fit curve $I(E)= (0.339\pm0.010) \times E_{{\rm keV}}^{-2}$}
   \label{Fig:lightcurve-ADAri}
   \end{center}
\end{figure}
Finally, we calculate the FHI (Fractional Halo Intensity) of
 Cygnus X-1 as a function of energy. Using the halo in 120 arcsec
instead of the whole halo to derive the FHI, we find
\begin{equation} I(E)= (0.339\pm0.010) \times E_{\rm{keV}}^{-2}
\end{equation}
as shown in Fig. 6. This is very different from the result derived
in Paper I. The main reason is that in Paper I we only used the
MARX 3.0 to calculate the PSF whose contribution is
under-estimated at large angles, especially in high energy bands.

Because for this observation of Cygnus X-3 (ObsID 425) only 512
bins of the CCDs were used, we can only reconstruct the halo as
much as 60 arcsec from the grating arm, the FHI of Cygnus X-3
cannot be calculated adequately.
\section{Conclusion and discussion}
\label{sect:conclusion} In this paper, we improve the method we
proposed in Paper I to resolve the point source halo from the {\it
Chandra} CC mode data and/or grating data. The method can resolve
the halo more accurately than the method in Paper I, even when
some data are contaminated. Taking advantage of the high angular
resolution of the {\it Chandra} instrument, we are able to probe
the intensity distribution of the X-ray halos as close as to 2-4
arcsec around their associated bright point sources.

 The FHI
derived from the Cygnus X-1 data is fitted by the $E^{-2}$ law
predicted theoretically, and also consistent with the results
obtained previously by Predehl $\&$ Schmit (1995) and Smith
\emph{et al}. (2002). Fitting the derived halo radial profile to
the WD01 model, we find the X-ray scattering dust clouds should be
located at between 1/2 to 1 of the distance to Cygnus X-1 and
between 1/6 to 3/4 (from model MRN) or 1/6 to 2/3 (from model
WD01) of the distance to Cygnus X-3, respectively. We also find
that the hydrogen column densities derived by fitting the WD01
have different values in different energy bands, as also noticed
by Smith \emph{et al}. (2002). This implies that our understanding
of the dust grain size distribution needs to be improved.

Although we have calculated PSF using the data from four
observations, the statistical uncertainties are still quite large
due to low counting rates of those sources, especially in the low
and high energy ends. It is expected that we should be able to
calculate the PSF more accurately using more data from point
sources with negligible halos in the futures. The limited
statistics of the radial halo profiles derived in this paper also
prevent us from probing the dust spatial distribution more
accurately; more and higher quality data are also needed for this
study.
\begin{acknowledgements}
J. Xiang thanks Yuxin Feng and Xiaoling Zhang for useful
discussions and insightful suggestions, and Dr Randall K. Smith
for providing the model code. This study is supported in part by
the Special Funds for Major State Basic Research Projects and by
the National Natural Science Foundation of China (project
no.10233030). Y. Yao acknowledges the support from NASA under the
contract NAS8-03060. SNZ also acknowledges supports by NASA's
Marshall Space Flight Center and through NASA's Long Term Space
Astrophysics Program, as well as the {\it Chandra} guest
investigation program.
\end{acknowledgements}

\label{lastpage}


\begin{thebibliography}{99}

  \bibitem[Marcos Raydan, Benar F. Svaiter]{iteration} Marcos Raydan, Benar F. Svaiter, 2002, Computational Optimization and Applications, 21,
  155
  \bibitem[Mathis, Rumpl $\&$ Nordsieck 1977]{mathis77} Mathis J. S., Rumpl,W., $\&$
  Nordsieck, K. H., 1977, ApJ, 217, 425
  \bibitem[Mathis $\&$ Lee 1991]{mathis91} Mathis J. S. $\&$ Lee
  C. W. 1991, ApJ, 376, 490
  \bibitem[Predehl, \textit{et al} 1995]{predehl95} Predehl, P. $\&$ Schmitt, J.H.M.M., 1995,
  A$\&$A, 293, 889
  \bibitem[Predehl, \textit{et al} 2000]{predehl00} Predehl, P. $\&$ Burwitz V., et al 2000,
  A$\&$A, 357, L25
  \bibitem[Smith, \textit{et al} 2002]{smith02} Smith R. K., $\&$
  Richard J. E., 2002, ApJ, 581, 562
  \bibitem[Trumper, \textit{et al} 1973]{Trump73} Trumper J., Schonfelder V., 1973,
  A$\&$A, 25, 445
  \bibitem[Weingartner $\&$ Draine (2001)]{wd01} Weingartner J.
  C., $\&$ Draine, B. T., 2001, ApJ, 548, 296
  \bibitem[Yao $\&$ Zhang \textit{et al} (2003)]{yao03} Yao, Y. S., Zhang, S. N., Zhang, X. L.
  $\&$ Feng, Y. X., 2003, ApJ, 594, L43


\end{thebibliography}
\end{document}